\documentclass[aps,prd,10pt,a4paper,groupedaddress,preprintnumbers,floatfix,nofootinbib,showpacs]{revtex4}
\usepackage[english]{babel}
\usepackage{amsmath}
\usepackage{graphicx}
\usepackage{color}
\usepackage{amssymb}
\usepackage{hyperref}
\usepackage{dcolumn}
\usepackage{bm}
\newcommand{\buno}{\mathbf{1}}
\newcommand{\bdos}{\mathbf{2}}
\newcommand{\btres}{\mathbf{3}}

\newcommand{\AddrAHEP}{%
  AHEP Group, Institut de F\'{\i}sica Corpuscular --
  C.S.I.C./Universitat de Val{\`e}ncia \\
  Edificio Institutos de Paterna, Apt 22085, E--46071 Valencia, Spain}

\begin{document}

\title{A new neutrino mass sum rule from  inverse seesaw}

\author{L.~Dorame}\email{dorame@ific.uv.es }\affiliation{$^{1}$\AddrAHEP}  
\author{S.~Morisi}\email{morisi@ific.uv.es}\affiliation{$^{1}$\AddrAHEP}
\author{E.~Peinado}\email{epeinado@ific.uv.es}\affiliation{$^{1}$\AddrAHEP}
\author{Alma D. Rojas} \email{alma.drp@gmail.com}\affiliation{$^{2}$ Facultad de Ciencias, CUICBAS,
Universidad de Colima, Colima, Mexico}
\author{J.~W.~F.~Valle} \email{valle@ific.uv.es}\affiliation{$^{1}$\AddrAHEP}



\date{\today}


\begin{abstract}

  A class of discrete flavor-symmetry-based models predicts
  constrained neutrino mass matrix schemes that lead to specific
  neutrino mass sum-rules (MSR).  One of these implies in a lower
  bound on the effective neutrinoless double beta mass parameter, even
  for normal hierarchy neutrinos. Here we propose a new model based on
  the $S_4$ flavor symmetry that leads to the new neutrino mass
  sum-rule and discuss how to generate a nonzero value for the reactor
  angle $\theta_{13}$ indicated by recent experiments, and the
  resulting correlation with the solar angle $\theta_{12}$.
 
\end{abstract}

\pacs{11.30.Hv,	
12.15.Ff, 
14.60.Pq 
}

\maketitle


\section{Introduction}

The discovery of neutrino oscillations have provided a strong evidence
of the non-vanishing neutrino masses, although their nature (if they
are Dirac or Majorana particles) has so far remained elusive. The
observation of neutrinoless double beta decay ($0\nu\beta\beta$) would
provide an irrefutable confirmation of the Majorana nature of
neutrinos~\cite{schechter:1982bd}.
Majorana neutrinos are characterized by a symmetric mass matrix whose
parameters are restricted by the experimental data: the neutrino
oscillation parameters, such as mixing angles and neutrino squared
mass differences~\cite{Schwetz:2011qt,schwetz:2008er}, as well as the
limits on the $0\nu\beta\beta$ effective mass
parameter~\cite{Rodejohann:2011mu,Barabash:2011fn}. Indeed, upcoming $0\nu\beta\beta$ experiments are expected to improve the sensitivity by up to about one order of magnitude~\cite{:2009yi,Abt:2004yk,Alessandrello:2002sj,GomezCadenas:2010gs}.

The general neutrino mixing matrix containing the three mixing angles
and the CP violating phases can be parametrized in different
equivalent
ways~\cite{schechter:1980gr,Rodejohann:2011vc,nakamura2010review}. A
particular \textit{ansatz} of the mixing matrix is the Tribimaximal
Mixing Matrix (TBM)~\cite{harrison:2002er} which, despite the fact of
the non-zero value of the $\theta_{13}$ angle indicated by recent
experiments~\cite{PhysRevLett.107.041801,Abe:2011fz}, can still be
used as a good first approximation. Specially so taking into account
that it can receive corrections from charged lepton diagonalization
and/or from renormalization effects, depending on its scale of
validity.

Several flavor models based in non-Abelian discrete symmetries predict
a two-parameter neutrino mass matrix which imply a particular mixing
matrix form, as is pointed out in \cite{harrison:2002er}.
In Ref.~\cite{Dorame:2011eb} it was noted that in these models only
the following mass relations can be obtained,
\begin{eqnarray}
 \chi m_2^{\nu}+\xi m_3^{\nu}=m_1^{\nu},\label{A}\\
 \frac{\chi}{m_2^{\nu}}+\frac{\xi}{m_3^{\nu}}=\frac{1}{m_1^{\nu}},\label{B}\\
{\chi}{\sqrt{m_2^{\nu}}}+{\xi}{\sqrt{m_3^{\nu}}}={\sqrt{m_1^{\nu}}},\label{C}\\
\frac{\chi}{\sqrt{m_2^{\nu}}}+\frac{\xi}{\sqrt{m_3^{\nu}}}=\frac{1}{\sqrt{m_1^{\nu}}},\label{D}
\end{eqnarray}
where $\chi$ and $\xi$ are free parameters that characterize each
specific model. A classification of all models predicting TBM mixing
which generate mass relations similar to the first three are given
there. The last case is completely new and here we will present a
model from first principles, implementing the inverse seesaw
mechanism~\cite{mohapatra:1986bd,gonzalez-garcia:1989rw} as well as a
non-Abelian flavor symmetry~\cite{Ishimori:2010au}, along the lines of
Ref.~\cite{Hirsch:2009mx}, but adopting $S_4$, instead of $A_4$.

\begin{figure}[h!]\label{seesawdiagram}
\includegraphics[width=0.40\textwidth]{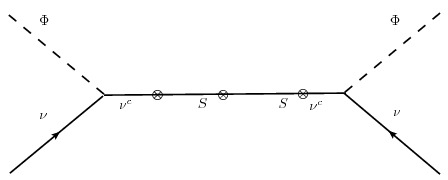}
\caption{Inverse seesaw mechanism.}
\end{figure}

The inverse seesaw scheme constitutes the first example of a low-scale
seesaw scheme~\cite{valle:2006vb} with naturally light neutrinos. 
The particle content is the same as that of the Standard Model
(SM) except for the addition of a pair of two component gauge singlet
leptons, $\nu_i$ and $S_i$, within each of the three generations,
labeled by $i$. The isodoublet neutrinos $\nu_i$ and the fermion
singlets $S_i$ have the same lepton number, opposite with respect to
that of the three singlets $\nu^c_i$ associated to the
``right-handed'' neutrinos.
In the $\nu$, $\nu^c$, $S$ basis the $9\times9$ neutral lepton mass
matrix $M_{\nu}$ has the form:
\begin{equation}
M_{\nu}= \left(\begin{array}{ccc}
       0& m_D^T& 0\\
       m_D& 0& M^T\\
      0& M& \mu
      \end{array}
 \right),\label{Mnu}
\end{equation}
where $Y_{\nu}$ and $M$ are arbitrary $3\times 3$ complex matrices,
while $\mu$ is symmetric due to the Pauli principle.  Following the
diagonalization seesaw method in \cite{Schechter:1981cv} one obtains
the effective light neutrino mass matrix as:
\begin{equation}
 m_{\nu}\sim m_D^T M^{T{-1}} \mu M^{-1} m_D,\label{seesawf}
\end{equation}
with the entry $\mu$ being very
small.  
The diagram illustrating the mass generation through the inverse seesaw
mechanism is shown in Figure~\ref{seesawdiagram}.

It is straightforward to show that if $m_D$ and $\mu$ are both
proportional to the identity, and
 \begin{equation}
M\sim M_{TBM}= \left(\begin{array}{ccc}
       x& y& y\\
       y&x+z& y-z\\
       y&y-z& x+z
      \end{array}
 \right), 
\end{equation}
in the basis where the charged lepton mass matrix is diagonal. Here
there is a specific (complex) relation among the parameters $x$, $y$
and $z$~\cite{Ma:2005qf}, leaving only two free complex parameters, we
obtain the sum mass rule in Eq.~(\ref{D}).


\vskip5.mm In the next section we give our model, in section III we
present the predictions regarding the lower bound on the neutrinoless
double beta decay amplitude, and discuss possible departures from
tribimaximality, including a finite $\theta_{13}$ value. In the
appendix~\ref{ScalarPotential} we present details on the symmetry
structure, Yukawa couplings as well as the scalar potential of the
model.

\section{The model}

Here we follow table I given in reference~\cite{Hirsch:2009mx}, where
some possibles schemes realizing the tri-bimaximal mixing pattern are
summarized for the inverse seesaw case. From these we will implement
case 1), since the other two cases, 2) and 3), correspond to the mass
sum rule relations C) and A) respectively, which have already model
realizations in the existing literature,
\begin{equation}
 M_D\propto \mathcal{I}\qquad \mu\propto \mathcal{I}, \qquad M \propto \left(\begin{array}{ccc}
       A & 0& 0\\
       0 & B&C\\
       0 & C&B
      \end{array}
 \right),
\end{equation}
In contrast to Ref.~\cite{Hirsch:2009mx} here we adopt the $S_4$
flavour symmetry, instead of $A_4$.

In order to obtain the $S_4$-based inverse seesaw model we assign the
charge matter fields as in table~\ref{tab1}. Three right handed
neutrinos $\nu_R$ are introduced, and three SU(2) fermion singlets
$S_i,i=1,2,3$, the latter transforming as the $\btres_1$ (note that
$\nu^c$ and $\nu_R$ are conjugates, hence have opposite lepton
number).  All fermion fields in table~\ref{tab1} transform as the
triplet $\btres_1$ and the Higgs doublet as the trivial singlet
$\buno_1$.

On the other hand, in order to generate the desired mass matrix
structures we introduce five flavon fields, $\phi_{\nu}$,
$\phi_{\nu}'$, $\phi_{l}$, $\phi_{l}'$, $\phi_{l}''$ supplemented by
the extra symmetries $Z_3$ and $Z_2$, whose assignments are given in
table~\ref{tabZ1}.
The presence of these extra Abelian symmetries in the theory ensures
the presence of adequate zeros in the neutrino mass matrix.
We keep renormalizability of the Lagrangian by adding a
Frogatt-Nielsen fermion $\chi$ and its conjugate $\chi^c $, both
singlets under the weak SU(2) gauge
group~\cite{Froggatt:1978nt,Barbieri:1996ww,deMedeirosVarzielas:2005ax,King:2006me}.
In table \ref{tabZ1} we present the relevant quantum numbers of the
matter fields in the theory under these extra symmetries.
 \begin{table}[h
]
\begin{center}
\begin{tabular}{|c||c|c|c|c||c|c|c|c|c|c|c||c|c|}
\hline
&$\overline{L}$&$\nu_{R}$&$l_{R}$&$h$&$S$&$\phi_{\nu}$&$\phi_{\nu}'$&$\phi_{l}$&$\phi_{l}'$&$\phi_{l}''$&$\sigma$&$\chi$&$\chi^c$\\
\hline
\hline
$SU(2)$&2&1&1&2&1&1&1&1&1&1&1&1&1\\
$S_4$&$3_{1}$&$3_{1}$&$3_{1}$&$1_{1}$&$3_{1}$&$3_{1}$&$1_{1}$&$3_{1}$&$3_{2}$&$1_{1}$&$1_{1}$&$3_{1}$&$3_{1}$\\
$U_l(1)$&-1&1&1&0&-1&0&0&0&0&0&2&1&-1\\
\hline
\end{tabular}
\caption{Fields and transformation properties under $SU(2)$, the $S_4$
  flavor symmetry, and global lepton number $U_l(1)$}\label{tab1}
\end{center}
\end{table}

\begin{figure}[ht]
  \includegraphics[width=0.50\textwidth,height=4.5cm]{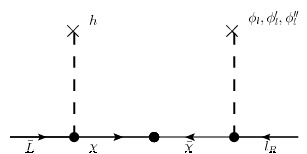}
\caption{Diagram  illustrating  charged lepton mass generation.}
\label{fig-ch}
\end{figure}

\begin{table}[h!]
\begin{center}
\begin{tabular}{|c||c|c|c|c||c|c|c|c|c|c|c||c|c|}
\hline
&$\overline{L}$&$\nu_{R}$&$l_{R}$&$h$&$S$&$\phi_{\nu}$&$\phi_{\nu}'$&$\phi_{l}$&$\phi_{l}'$&$\phi_{l}''$&$\sigma$&$\chi$&$\chi^c$\\
\hline
\hline
$Z_3$&$\omega^2$&$\omega$&1&1&1&$\omega^2$&$\omega^2$&$\omega$&$\omega$&$\omega$&1&$\omega$&$\omega^2$\\
$Z_2$&+&+&+&+&-&-&-&+&+&+&+&+&+\\
\hline
\end{tabular}
\caption{Fields and their transformation properties under the $Z_3$,
  and $Z_2$ flavor symmetries}\label{tabZ1}
\end{center}
\end{table}
The renormalizable Lagrangian relevant for neutrinos is
\begin{equation}
\mathcal{L}_{\nu} = Y_{D_{ij}}\overline{L}_i\nu_{R_j}h+Y^{k}_{\nu ij}\nu_{R_i}S_j\phi_{\nu_{k}}+   Y'_{\nu ij}\nu_{R_i}S_j\phi_{\nu}' + \mu_{ij}S_iS_j\sigma,
\end{equation}
while the renormalizable Yukawa terms involving the messenger fields is 
\begin{equation}\label{RenChPot}
  \mathcal{L}_{\chi}= M_{\chi}\chi\chi^c+\bar{L}h\chi+\chi^c l_R\Phi_l+\chi^c l_R\Phi_l'+\chi^c l_R\Phi_l''.
\end{equation}
After integrating out the messenger fields $\chi$, the effective
Lagrangian for charged leptons takes the form
\begin{equation}\label{EffLag}
 \mathcal{L}_{l}=\frac{y_l}{\Lambda}(\bar{L}l_R)h \phi_{l}+\frac{y_l'}{\Lambda}(\bar{L}l_R)h \phi_{l}'+\frac{y_l''}{\Lambda}(\bar{L}l_R)h \phi_{l}'',
\end{equation}
where $\Lambda$ is the effective scale.  This effective Lagrangian is
responsible for charged lepton mass generation, as shown in figure
\ref{fig-ch}.

In order to obtain the desired neutrino mixing matrix we require the
flavon fields to have the following alignments:
\begin{equation}
\label{eq:align}
\langle\phi_{\nu}\rangle =v_{\nu} (1,0,0), \quad
\langle\phi_{l}\rangle =v_{l} (1,1,1), \quad
\langle\phi_{l'}\rangle = v_{l'}(1,1,1),
\end{equation}
where we also define $\langle\phi_{\nu'}\rangle = v_{\nu'}$,
$\langle\phi_{l''}\rangle = v_{l''} $, $\langle \sigma \rangle =
v_{\sigma} $ and $\langle h\rangle= v$.  In
appendix~\ref{ScalarPotential} we report the form of the potential
(\ref{potential}).  We have verified that there exists a large portion
of the parameter space where the required alignment is found to be a
solution of minimization of the potential.

With these alignments the three $3\times3$ blocks in Eq.~(\ref{Mnu})
and charged lepton matrices take the form
\begin{equation}
(\mu)= \left(\begin{array}{ccc}
       \mu v_{\sigma}& 0& 0\\
      0& \mu v_{\sigma}&0\\
       0& 0&\mu v_{\sigma}
      \end{array}
 \right),
\quad
M_D = \left(\begin{array}{ccc}
       Y_D v&  0& 0\\
      0& Y_Dv&0\\
       0& 0&Y_D v
      \end{array}
 \right),
\quad
M = \left(\begin{array}{ccc}
      Y_{\nu'} v_{\nu'}&  0& 0\\
      0&  Y_{\nu '}v_{\nu'}&Y_{\nu}v_{\nu} \\
       0&  Y_{\nu}v_{\nu}& Y_{\nu '}v_{\nu'}
      \end{array}
 \right),
\end{equation}
and
\begin{equation}
M_l = \left(\begin{array}{ccc}
       y_{l ''} v_{l''} & y_{l}v_{l} - y_{l '} v_{l'} &  y_{l}v_{l} +y_{l'}v_{l'} \\
       y_{l}v_{l} + y_{l'}v_{l'}  & y_{l''}v_{l''} & y_{l}v_{l} -y_{l'}v_{l'} \\
      y_{l}v_{l} - y_{l'}v_{l'} & y_{l}v_{l} + y_{l'}v_{l'} & y_{l''}v_{l''} 
      \end{array}
 \right) \frac{v}{\Lambda}.\label{chargedmass} 
\end{equation}
The charged lepton mass matrix, Eq. (\ref{chargedmass}), is
diagonalized by the "magic" matrix
\begin{equation}
U_{\omega}=\frac{1}{\sqrt{3}} \left(\begin{array}{ccc}
       1& 1& 1\\
      1&  \omega&\omega^2\\
       1& \omega^2&\omega
      \end{array}
 \right).
\end{equation}
On the other hand, by using Eqs.~(\ref{Mnu}) and (\ref{seesawf}) it is
straightforward to obtain the light neutrino mass matrix which takes
the form
\begin{equation}
M_{\nu}=
\left(
\begin{array}{ccc}
 \frac{1}{a^2} & 0 & 0 \\
 0 & \frac{a^2+b^2}{\left(b^2-a^2\right)^2} & -\frac{2 a b}{\left(b^2-a^2\right)^2} \\
 0 & -\frac{2 a b}{\left(b^2-a^2\right)^2} & \frac{a^2+b^2}{\left(b^2-a^2\right)^2} 
\end{array}
\right)
\end{equation}
where $a=Y_{\nu'}v_{\nu'}/(\sqrt{\mu v_{\sigma}} Y_D v)$ and
$b=Y_{\nu}v_{\nu}/(\sqrt{\mu v_{\sigma}} Y_D v)$.  In the basis where
charged lepton mass matrix is diagonal, the light neutrino mass matrix
is diagonalized by the TBM-form, and the corresponding eigenvalues are
given by
\begin{equation}
\begin{array}{l}
m_1=\frac{1}{(a+b)^2},\\
m_2=\frac{1}{(a-b)^2},\\
m_3=\frac{1}{a^2}.\\
\end{array}
\end{equation}
With these eigenvalues we obtain the neutrino mass sum rule
\begin{equation}
\frac{1}{\sqrt{m_1}}=\frac{2}{\sqrt{m_3}}-\frac{1}{\sqrt{m_2}}.
\end{equation}
which is, indeed, of the type given in Eq.~(\ref{D}). 

\section{Phenomenology}

\subsection{Neutrinoless double beta decay}

Using the symmetric parametrization of the lepton mixing
matrix~\cite{schechter:1980gr,Rodejohann:2011vc} we can obtain the
general expression of the mass parameter $|m_{ee}|$ which determines
the $0\nu\beta\beta$ decay amplitude as

\begin{equation} \label{eq:meff_expl}
 |m_{ee}| = \left|\sum_j U_{ej}^2 \, m_j \right| 
= 
\left\{ 
\begin{array}{cc} \left| c_{12}^2 c_{13}^2  \,m_1 \, 
\text{e}^{2 i \alpha} + s_{12}^2 c_{13}^2 \, m_2 \, \text{e}^{2 i \beta}+
s_{13}^2  \,m_3 \,\text{e}^{2 i \delta} \right| 
& \mbox{(PDG)} \, , \\ 
\left| c_{12}^2 c_{13}^2  \,m_1 + s_{12}^2 c_{13}^2  \,m_2 \,\text{e}^{2 i \phi_{12}}+
s_{13}^2  \, m_3  \,\text{e}^{2 i \phi_{13}}
\right| & \mbox{(symmetrical)} \, . 
 \end{array} \right.    
  \end{equation}
  where $c_{ij}=\cos\theta_{ij}$ and $s_{ij}=\sin\theta_{ij}$,
  $m_i,i=1,2,3$ the neutrino masses, and we adopt the symmetric
  parametrization where $ \phi_{12}$ and $ \phi_{13}$ are the two
  Majorana phases.

  By varying the neutrino oscillation parameters in their allowed
  range one can plot $|m_{ee}|$ in terms of the lightest neutrino
  mass. 
  Depending on which is the lightest neutrino one can have two
  different spectra, normal and inverse hierarchy, respectively. In
  the latter case one has a lower bound, on quite general grounds, as
  in this case there can be no destructive interference between the
  light neutrinos.

  In the present scheme, however, as noted in
  Ref.~\cite{Dorame:2011eb}, the neutrino mass sum-rule can be
  interpreted geometrically as a triangle in the complex plane, its
  area providing a measure of the Majorana CP violation. Then, fixing
  the $(\xi,\chi)$ parameters for each model one can, in principle,
  determine the two Majorana CP violating phases. 

\begin{figure}[!h]
\includegraphics[width=0.7\textwidth]{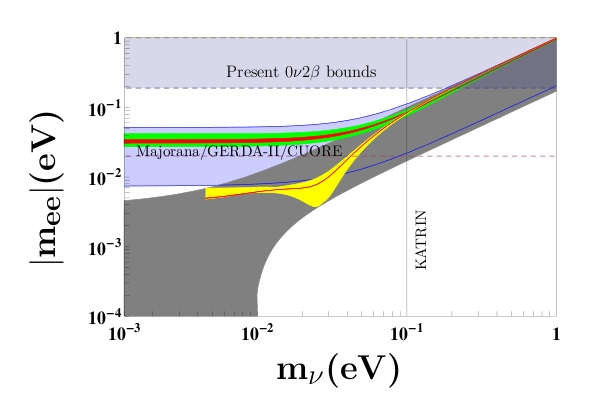} 
\caption{$|m_{ee}|$ as a function of the lightest neutrino mass
  corresponding to the mass sum-rule in Eq.~(\ref{D}). The grey and
  dark bands correspond to generic normal and inverse hierarchy
  regions, while the green and yellow bands correspond to our flavour
  prediction varying the values of oscillation parameters in their
  3~$\sigma$ C.L. range. The thin red bands correspond to the TBM
  limit. For references to the experiments see~\cite{:2009yi,Abt:2004yk,Alessandrello:2002sj,GomezCadenas:2010gs,Osipowicz:2001sq}}
\label{fig:dbs}
\end{figure}
As a result there is a lower bound on $|m_{ee}|$ even in the case of
normal hierarchy (for other schemes of this type see, for example,
Ref.~\cite{Dorame:2011eb} and references therein), as illustrated in
Fig.~\ref{fig:dbs}.

\subsection{Quark sector}

Quarks are introduced as in table\,\ref{tabQ} where, differently from the lepton
sector, we assign the fist and second families to a doublet representation of $S_4$
and the third family to a singlet of $S_4$, namely $Q_D=(Q_1,Q_2)\sim 2$,
$q_{R_D}=(q_{R_1},q_{R_2})\sim 2$, $Q_s=Q_3\sim 1_1$  and $=q_{R_3}\sim 1_1$. 
\begin{table}[h!]
\begin{center}
\begin{tabular}{|c||c|c|c|c|c|c|c|c|}
\hline
&$\overline{Q}_D$&$\overline{Q}_S$&$u_{R_D}$&$u_{R_S}$&$d_{R_D}$&$d_{R_S}$&$\phi_D$&$\phi_S$\\
\hline
\hline
$SU(2)$&2&2&1&1&1&1&1&1\\
$S_4$&2&$1_1$&2&$1_1$&2&$1_1$&2&$1_1$\\
$Z_3$&$\omega$&$\omega$&$\omega^2$&$\omega^2$&$\omega$&$\omega$&$\omega$&$\omega$\\
$Z_2$&+&+&-&-&-&-&-&-\\
\hline
\end{tabular}
\end{center}
\caption{Quark sector and their transformation properties under the $Z_3$,
  and $Z_2$ flavor symmetries}\label{tabQ}
\end{table}
We add flavons $\phi_{D,S}$ in doublet and singlet representations of
the $S_4$,
\begin{equation}
  \mathcal{L}_{q}^{d}=(Y_1^{d} \overline{Q}_S d_{R_{S}} \phi_S + Y_2^{d} \overline{Q}_D \phi_D d_{R_D} + Y_3^{d}\overline{Q}_D  d_{R_D} \phi_S + Y_4^{d} \overline{Q}_D \phi_D d_{R_S} + Y_5^{d} \overline{Q}_S \phi_D d_{R_D} )h/\Lambda+ h.c.,
\end{equation}
\begin{equation}
 \mathcal{L}_{q}^{u}=(Y_1^{u} \overline{Q}_S u_{R_{S}} \tilde{\phi}_S + Y_2^{u} \overline{Q}_D \tilde{\phi}_D u_{R_D} + Y_3^{u}\overline{Q}_D  u_{R_D} \tilde{\phi}_S + Y_4^{u} \overline{Q}_D \tilde{\phi}_D u_{R_S} + Y_5^{u} \overline{Q}_S \tilde{\phi}_D u_{R_D})\tilde{h}/\Lambda + h.c.,
\end{equation}
As in the charged lepton sector the dimension five operators can be
given in terms of renormalizable interaction by introducing suitable
messenger fields.  Taking the VEV of $\phi_D$ in the direction (we
verified that it is a possible solution of the potential)
\begin{equation}
\langle\phi_{D}\rangle \sim (-\sqrt{3},1) \ ,
\end{equation}
the mass matrix for the quarks is 
\begin{equation}
\textbf{M}_{u(d)} = \left(\begin{array}{ccc}
       m_1^{u(d)}+ m_2^{u(d)}&-\sqrt{3} \ m_2^{u(d)}& -\sqrt{3} \ m_5^{u(d)}\\
      -\sqrt{3} \ m_2^{u(d)}& m_1^{u(d)} - m_2^{u(d)}&m_5^{u(d)}\\
       -\sqrt{3} \ m_4^{u(d)}& m_4^{u(d)}&m_3^{u(d)}
      \end{array}
\right),
\end{equation}
which is very similar to the one proposed in Ref.\,\cite{Kubo:2003iw}
where a fit of the quarks masses and mixing has been performed and we
refer to that paper for more detail.

\subsection{Finite $\theta_{13}$ value}

As we have discussed so far the model leads to the tribimaximal mixing
pattern. However by coupling an extra $S_4$-doublet flavon field one
can obtain corrections from the charged lepton sector which induce
nonzero values of $\theta_{13}$ as recently suggested by the
T2K~\cite{PhysRevLett.107.041801} and Double-Chooz~\cite{Abe:2011fz}
results \cite{Schwetz:2011qt} including also recent reactor flux
calculations. 

\begin{figure}[!h]
\includegraphics[width=0.45\textwidth]{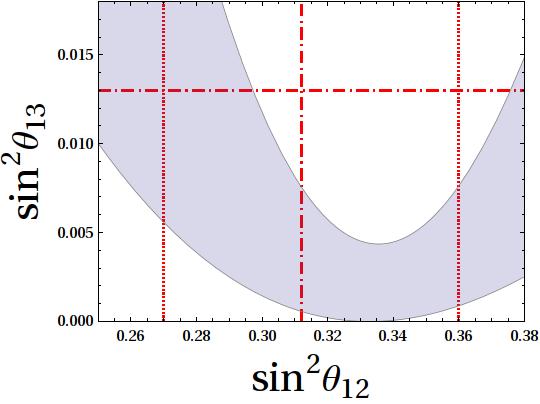} 
\caption{Correlations between reactor and solar neutrino mixing
  angles. See text for explanations.}
\label{fig:corr}
\end{figure}

For example, consider a flavon scalar doublet under $S_4$, $\Phi \sim
\bdos$ and transforming as $(\omega,+)$ under $Z_3\times Z_2$.  In the
Lagrangian we must then include the term
\begin{equation}
(\bar{L}l_R)h \Phi.
\end{equation}
This is a dimension five operator which can be obtained from a
renormalizable Lagrangian by means of the messenger fields
$\chi,\chi^c$ of table~\ref{tab1} as shown in figure~(\ref{fig-ch}).
Assuming that $\Phi$ acquires VEV $\left\langle \Phi\right\rangle
=(u_1,u_2)$, a natural vacuum alignment is $u_1=-\sqrt{3}u_2$, since
this is consistent with the previous alignments in Eq.~(\ref{eq:align}).
Using multiplication rules in Appendix A one finds that the
contribution from this term to the charged lepton mass matrix is
\begin{equation}
\begin{array}{lll}
  \delta M_l&= &\left(\begin{array}{ccc}
       -\sqrt{\frac{2}{3}}v u_2& 0& 0\\
      0& \sqrt{\frac{1}{2}}v u_1+\sqrt{\frac{1}{6}}v u_2 & 0\\
       0& 0&-\sqrt{\frac{1}{2}}v u_1+\sqrt{\frac{1}{6}}v u_2 
      \end{array}
 \right)
\end{array}
\end{equation}
which modifies the diagonal entries $\delta M_l$ in the charged lepton
mass matrix, $M_l$, so that the total $M_l + \delta M_l$ is no longer
diagonalized by $U_\omega$.  This way one can induce a potentially
``large'' value for $\theta_{13}$, as hinted by recent
experiments~\cite{PhysRevLett.107.041801,Abe:2011fz}, and also
potential departures of the solar and atmospheric angles from their
TBM values.
Moreover, in the presence of a nonzero $\theta_{13}$ one finds
correlations among these neutrino mixing angles. The most interesting
of these involves the solar angle, as illustrated in
Fig.~\ref{fig:corr}.
The horizontal dot-dashed line indicates the best fit in the global
analysis in Ref.~\cite{Schwetz:2011qt}.
On the other hand the region in the vertical band delimited by the
dotted lines corresponds to the 3$\sigma$ region found in
Ref.~\cite{Schwetz:2011qt}.
One sees how the departure of the solar $\theta_{12}$ from its TBM
value can be substantial and large$\theta_{13}$ values require the
solar angle to lie \texttt{below} the TBM prediction~\footnote{ For
  technical simplicity we have varied the charged lepton massess well
  above what is allowed by their current determinations. The real
  correlation is expressed as a subband of this}.
A nonzero $\theta_{13}$ would also open the way also for the
phenomenon of CP violation in neutrino oscillations, one of the
central goals of the upcoming generation of long baseline oscillation
studies~\cite{Nunokawa:2007qh,Bandyopadhyay:2007kx}.

\begin{acknowledgments}

  This work was supported by the Spanish MEC under grants
  FPA2011-22975 and MULTIDARK CSD2009-00064 (Consolider-Ingenio 2010
  Programme), by Prometeo/2009/091 (Generalitat Valenciana), by the EU
  ITN UNILHC PITN-GA-2009-237920. S. M. is supported by a Juan de la
  Cierva contract. E. P. is supported by CONACyT (Mexico).
  A. D. R. is supported by Fundaci\'on Carolina and would like to
  thank hospitality at IFIC while this work was carried out and
  Alfredo Aranda for his helpful suggestions.  L. D. is supported by a
  CSIC JAE predoctoral fellowship.

\end{acknowledgments}

\appendix

\section{$S_4$ group}

The $S_4$ group is the discrete group given by the four objects permutations. It contains 24 elements and can be obtained from two generators, $S$ and $T$, satisfying:
\begin{equation}
 S^4=T^3=1,\qquad S T^2 S=T.
\end{equation}
 
The $S_4$ irreducible representations are two singlets, $\mathbf{1}_1, \mathbf{1}_2$, one doublet, $\bdos$, and two triplets, $\btres_1, \btres_2$. The product rules are given by (for more details see  \cite{Hagedorn:2006ug}.)
 
\begin{equation}
\begin{array}{l}
\buno_i\times\buno_j=\buno_{(i+j)\mbox{ \tiny{mod2 +1}}} \; \forall\;i,j\\
\bdos\times\buno_i=2 \;\forall\;i\\
\btres_i\times\buno_j=\buno_{(i+j)\mbox{ \tiny{mod2 +1}}} \; \forall\;i,j\\
\btres_i\times\bdos=\btres_1+\btres_2 \;\forall\;i\\
\btres_1\times\btres_2=\buno_2+\bdos+\btres_1+\btres_2 \\
\bdos\times\bdos=\buno_1+\bdos+\buno_2\\
\btres_i\times\btres_i=\buno_1+\bdos+\btres_1+\btres_2 \;\forall\;i
\end{array}
\end{equation}
where we can introduce the notation $[\mu\times\mu]$ and $\left\lbrace \mu\times\mu\right\rbrace$ for the symmetric and antisymmetric part of $\mu\times\mu$ respectively:

\begin{equation}
\begin{array}{l}
 [\bdos\times\bdos]=\buno_1+\bdos,\quad  \left\lbrace \bdos\times\bdos\right\rbrace=\buno_2\\
\left[ \btres_i\times\btres_i\right] = \buno_1+\bdos+\btres_1,\quad  \left\lbrace \btres_i\times\btres_i\right\rbrace=\btres_2 \;\forall i
\end{array}.
\end{equation}
Given the following representations 
\begin{equation}
 \begin{array}{l}
  A,A'\sim \buno_1, B,B'\sim \buno_2,\;
 \left( \begin{array}{c}
a_1\\
a_2
\end{array}
\right),
 \left( \begin{array}{c}
a_1'\\
a_2'
\end{array}
\right)\sim 2,\;
\left( \begin{array}{c}
b_1\\
b_2\\
b_3
\end{array}
\right),
\left( \begin{array}{c}
b_1'\\
b_2'\\
b_3'
\end{array}
\right)\sim \btres_1\;
\left( \begin{array}{c}
c_1\\
c_2\\
c_3
\end{array}
\right),
\left( \begin{array}{c}
c_1'\\
c_2'\\
c_3'
\end{array}
\right)\sim \btres_2.          
 \end{array}
\end{equation}

As the representation matrices can be choosen all reals, the conjugate representations transform in the same way.

For the product of one dimensional representations the Clebsch Gordan coefficients are the trivial products of representations, and also for the product of the $1_1$ singlet with any non-trivial representation
\begin{equation}
 \left( \begin{array}{c}
A a_1\\
A a_2
\end{array}
\right)\sim 2,\;
\left( \begin{array}{c}
A b_1\\
A b_2\\
A b_3
\end{array}
\right)\sim\btres_1,\;
\left( \begin{array}{c}
A c_1\\
A c_2\\
A c_3
\end{array}
\right)\sim\btres_2.
\end{equation}
For the product with the $1_2$ singlet
\begin{equation}
 \left( \begin{array}{c}
-B a_2\\
B a_1
\end{array}
\right)\sim 2,\;
\left( \begin{array}{c}
A b_1\\
A b_2\\
A b_3
\end{array}
\right)\sim\btres_2,\;
\left( \begin{array}{c}
A c_1\\
A c_2\\
A c_3
\end{array}
\right)\sim\btres_1.
\end{equation}
The Clebsch Gordan coefficients for the product $\bdos\times\bdos$ are
\begin{equation}
 \begin{array}{l}
  a_1 a_1'+a_2 a_2'\sim \buno_1\\
 -a_1 a_2'+a_2 a_1'\sim \buno_2\\
   \left( \begin{array}{c}
a_1 a_2'+a_2 a_1'\\
 a_1 a_1'-a_2 a_2'
\end{array}
\right)\sim 2,
 \end{array}
\end{equation}
for $\btres_1\times\btres_1$
\begin{equation}
 \begin{array}{l}
 \sum_{j=1} ^3 b_j b_j'\sim \buno_1,\\
 \left( \begin{array}{c}
\frac{1}{\sqrt{2}}(b_2 b_2'-b_3 b_3')\\
\frac{1}{\sqrt{6}}(-2b_1 b_1'+b_2 b_2'+b_3 b_3')
\end{array}
\right)\sim 2,\\
 \left( \begin{array}{c}
b_2 b_3'+b_3 b_2'\\
b_1 b_3'+b_3 b_1'\\
b_1 b_2'+b_2 b_1'
\end{array}
\right)\sim\btres_1, \qquad
\left( \begin{array}{c}
b_3 b_2'-b_2 b_3'\\
b_1 b_3'-b_3 b_1'\\
b_2 b_1'-b_1 b_2'
\end{array}
\right)\sim\btres_2,
 \end{array}
\end{equation}
for $\btres_2\times\btres_2$
\begin{equation}
 \begin{array}{l}
 \sum_{j=1} ^3 c_j c_j'\sim \buno_1,\\
 \left( \begin{array}{c}
\frac{1}{\sqrt{2}}(c_2 c_2'-c_3 c_3')\\
\frac{1}{\sqrt{6}}(-2c_1 c_1'+c_2 c_2'+c_3 c_3')
\end{array}
\right)\sim 2,\\
 \left( \begin{array}{c}
c_2 c_3'+c_3 c_2'\\
c_1 c_3'+c_3 c_1'\\
c_1 c_2'+c_2 c_1'
\end{array}
\right)\sim\btres_1, \qquad
\left( \begin{array}{c}
c_3 c_2'-c_2 c_3'\\
c_1 c_3'-c_3 c_1'\\
c_2 c_1'-c_1 c_2'
\end{array}
\right)\sim\btres_2.
 \end{array}
\end{equation}
For the couplings $\bdos\times\btres_1$ and $\bdos\times\btres_2$, we have respectively
\begin{equation}
 \begin{array}{ll}
  \left( \begin{array}{c}
a_2 b_1\\
-\frac{1}{2}(\sqrt{3}a_1 b_2+ a_2 b_2)\\
\frac{1}{2}(\sqrt{3}a_1 b_3 - a_2 b_3)
\end{array}
\right)\sim\btres_1, & \left(\begin{array}{c}
a_1 c_1\\
\frac{1}{2}(\sqrt{3}a_2 c_2 - a_1 c_2)\\
-\frac{1}{2}(\sqrt{3}a_2 c_3 + a_1 c_3)
\end{array}
\right)\sim\btres_1 , \\
\left( \begin{array}{c}
a_1 b_1\\
\frac{1}{2}(\sqrt{3}a_2 b_2 - a_1 b_2)\\
-\frac{1}{2}(\sqrt{3}a_2 b_3 + a_1 b_3)
\end{array}
\right)\sim\btres_2, &  \left(\begin{array}{c}
a_2 c_1\\
-\frac{1}{2}(\sqrt{3}a_1 c_2 + a_2 c_2)\\
\frac{1}{2}(\sqrt{3}a_1 c_3 - a_2 c_3)
\end{array}
\right)\sim\btres_2.
 \end{array}
\end{equation}
And finally, for the $\btres_1\times\btres_2$  product 
\begin{equation}
 \begin{array}{l}
 \sum_{j=1} ^3 b_j c_j\sim \buno_2,\\
 \left( \begin{array}{c}
\frac{1}{\sqrt{6}}(2b_1 c_1-b_2 c_2-b_3 c_3)\\
\frac{1}{\sqrt{2}}(b_2 c_2-b_3 c_3)
\end{array}
\right)\sim 2,\\
\left( \begin{array}{c}
b_3 c_2-b_2 c_3\\
b_1 c_3-b_3 c_1\\
b_2 c_1-b_1 c_2
\end{array}
\right)\sim\btres_1, \qquad
 \left( \begin{array}{c}
b_2 c_3+b_3 c_2\\
b_1 c_3+b_3 c_1\\
b_1 c_2+b_2 c_1
\end{array}
\right)\sim\btres_2.
 \end{array}
\end{equation}

\section{Yukawa couplings}
Each term in the Lagrangian for neutrinos and charged leptons can be decomposed by components as:
\begin{equation}
 \mu S\cdot S \sigma= \mu (S_1 S_1+S_2 S_2+S_3 S_3)\sigma,
\end{equation}
\begin{equation}
 Y_D \bar{L}\cdot \nu_R h= Y_D(\bar{L}_1 \nu_{1R}+\bar{L}_2 \nu_{2R}+\bar{L}_3 \nu_{3R})h,
\end{equation}
\begin{equation}
 Y_{\nu} (\nu_R\cdot S)\phi_{\nu}= Y_{\nu}[(\nu_{2R} S_{3R}+\nu_{3R} S_2)\phi_{\nu 1}+(\nu_{1R} S_{3R}+\nu_{3R} S_1)\phi_{\nu 2}+(\nu_{1R} S_{2R}+\nu_{2R} S_1)\phi_{\nu 3}],
\end{equation}
\begin{equation}
 Y_{\nu '} (\nu_R\cdot S)\phi_{\nu '}= Y_{\nu '}(\nu_{1R} S_1+\nu_{2R}S_2+\nu_{3R}S_3)\phi_{\nu '},
\end{equation}

\begin{equation}
\frac{y_l}{\Lambda}(\bar{L}l_R)h \phi_{l}=\frac{y_l}{\Lambda}[(\bar{L}_2l_{3R}+\bar{L}_3l_{2R})h  \phi_{l1}+(\bar{L}_1 l_{3R}+\bar{L}_3 l_{1R})h  \phi_{l2}+(\bar{L}_1 l_{2R}+\bar{L}_2 l_{1R})h  \phi_{l3}]
\end{equation}

\begin{equation}
\frac{y_l'}{\Lambda}(\bar{L}l_R)h \phi_{l'}=\frac{y_l'}{\Lambda}[(\bar{L}_3l_{2R}-\bar{L}_2l_{3R})h  \phi_{l'1}+(\bar{L}_1l_{3R}-\bar{L}_3l_{1R})h  \phi_{l'2}+(\bar{L}_2l_{1R}-\bar{L}_1l_{2R})h  \phi_{l'3}]
\end{equation}

\begin{equation}
 \frac{y_l''}{\Lambda}(\bar{L}l_R)h \phi_{l''}=\frac{y_l''}{\Lambda}(\bar{L}_1 l_{1R}+\bar{L}_2l_{2R}+\bar{L}_3l_{3R})h \phi_{l''}]
\end{equation}

\section{Scalar potential}
\label{ScalarPotential}
The most general renormalizable scalar potential is (without write the $S_4$ products explicitly):
\begin{eqnarray}
   V &=&V(h)+V(\sigma)+V(\Phi_{\nu})+V(\Phi_{\nu'})+V(\Phi_{l})+V(\Phi_{l'})+V(\Phi_{l''})\\
&&+V(\Phi_{\nu},\Phi_{\nu'},\Phi_{l},\Phi_{l'},\Phi_{l''})+V(\sigma,h,\Phi_{\nu},\Phi_{\nu'},\Phi_{l},\Phi_{l'},\Phi_{l''}),\label{potential}
\end{eqnarray}
with
\begin{eqnarray*}
 V(h)&=&\mu_h h^{\dag}h+\lambda_h(h^{\dag}h)(h^{\dag}h),\\
V(\sigma)&=&\mu_{\sigma} \sigma^{\dag}\sigma+\lambda_{\sigma}(\sigma^{\dag}\sigma)(\sigma^{\dag}\sigma)\\
V(\Phi_{\nu})&=&\mu_1(\Phi_{\nu}^{\dag}\Phi_{\nu})+\sum_{i}\lambda_{j}^{\nu}\left\lbrace \Phi_{\nu}^{\dag}\Phi_{\nu}\Phi_{\nu}^{\dag}\Phi_{\nu}\right\rbrace_i  \\
V(\Phi_{\nu'}))&=&\mu_2(\Phi_{\nu'}^{\dag}\Phi_{\nu'})+\sum_{i}\lambda_{j}^{\nu'}\left\lbrace \Phi_{\nu'}^{\dag}\Phi_{\nu'}\Phi_{\nu'}^{\dag}\Phi_{\nu'}\right\rbrace_i  \\
V(\Phi_{l})&=&\mu_3(\Phi_{l}^{\dag}\Phi_{l})+\sum_{i}\lambda_{i}^{l}\left\lbrace \Phi_{l}^{\dag}\Phi_{l}\Phi_{l}^{\dag}\Phi_{l}\right\rbrace_i +\sum_i\kappa_i\left\lbrace  (\Phi_{l}\Phi_{l})\Phi_{l}+ h.c.\right\rbrace_i, \\
V(\Phi_{l'})&=&\mu_4(\Phi_{l'}^{\dag}\Phi_{l'})+\sum_{i}\lambda_{i}^{l'}\left\lbrace \Phi_{l'}^{\dag}\Phi_{l'}\Phi_{l'}^{\dag}\Phi_{l'}\right\rbrace_i +\sum_i\kappa_i\left\lbrace  (\Phi_{l'}\Phi_{l'})\Phi_{l'}+ h.c.\right\rbrace_i, \\
V(\Phi_{l''})&=&\mu_5(\Phi_{l''}^{\dag}\Phi_{l''})+\sum_{i}\lambda_{i}^{l''}\left\lbrace \Phi_{l''}^{\dag}\Phi_{l''}\Phi_{l''}^{\dag}\Phi_{l''}\right\rbrace_i +h.c., \\
V(\sigma,h,\Phi_{\nu},\Phi_{\nu'},\Phi_{l},\Phi_{l'},\Phi_{l''})&=&\lambda^{h\sigma}(h^{\dag}h)(\sigma^{\dag}\sigma)+\lambda^{\nu\sigma}(\Phi_{\nu}^{\dag}\Phi_{\nu})(\sigma^{\dag}\sigma)+\lambda^{\nu'\sigma}(\Phi_{\nu'}^{\dag}\Phi_{\nu'})(\sigma^{\dag}\sigma)\\
&&+\lambda^{l\sigma}(\Phi_{l}^{\dag}\Phi_{l})(\sigma^{\dag}\sigma)+\lambda^{l'\sigma}(\Phi_{l'}^{\dag}\Phi_{l'})(\sigma^{\dag}\sigma)+\lambda^{l''\sigma}(\Phi_{l''}^{\dag}\Phi_{l''})(\sigma^{\dag}\sigma),\\
\end{eqnarray*}
\begin{eqnarray*}
V(\Phi_{\nu},\Phi_{\nu'},\Phi_{l},\Phi_{l'},\Phi_{l''})&=&\sum_i \kappa_i\left\lbrace  \Phi_{l}\Phi_{l}\Phi_{l'}\right\rbrace_i+\sum_i \kappa_i\left\lbrace  \Phi_{l}\Phi_{l'}\Phi_{l'}\right\rbrace_i\\
&&+\sum_i \kappa_i\left\lbrace  \Phi_{l}\Phi_{l}\Phi_{l''}\right\rbrace_i+\sum_i \kappa_i\left\lbrace  \Phi_{l'}\Phi_{l'}\Phi_{l''}\right\rbrace_i \\
&&+\sum_i \kappa_i\left\lbrace\Phi_{l}^{\dag}\Phi_{\nu}\Phi_{\nu}\right\rbrace_i+\sum_i \kappa_i\left\lbrace  \Phi_{l'}^{\dag}\Phi_{\nu}\Phi_{\nu}\right\rbrace_i+\sum_i \kappa_i\left\lbrace  \Phi_{l''}^{\dag}\Phi_{\nu}\Phi_{\nu}\right\rbrace_i \\
&&+\sum_i \kappa_i\left\lbrace  \Phi_{l}^{\dag}\Phi_{\nu}\Phi_{\nu'}\right\rbrace_i +\sum_i \kappa_i\left\lbrace  \Phi_{l''}^{\dag}\Phi_{\nu'}\Phi_{\nu'}\right\rbrace_i \\
&&+\sum_{i}\lambda_{i}^{lll'l'}\left\lbrace \Phi_{l}^{\dag}\Phi_{l}^{\dag}\Phi_{l'}\Phi_{l'}\right\rbrace_i+\sum_{i}\lambda_{i}^{ll'l'l'}\left\lbrace \Phi_{l}\Phi_{l'}^{\dag}\Phi_{l'}\Phi_{l'}\right\rbrace_i \\
&&+\sum_{i}\lambda_{i}^{llll'}\left\lbrace \Phi_{l}^{\dag}\Phi_{l}^{\dag}\Phi_{l}\Phi_{l'}\right\rbrace_i+\sum_{i}\lambda_{i}^{llll'}\left\lbrace \Phi_{l'}^{\dag}\Phi_{l}^{\dag}\Phi_{l}\Phi_{l'}\right\rbrace_i\\
&&+\sum_{i}\lambda_{i}^{ll'l'l'}\left\lbrace \Phi_{l}^{\dag}\Phi_{l'}^{\dag}\Phi_{l'}\Phi_{l'}\right\rbrace_i+\sum_{i}\lambda_{i}^{llll'}\left\lbrace \Phi_{l}^{\dag}\Phi_{l}^{\dag}\Phi_{l}\Phi_{l''}\right\rbrace_i\\
&&+\sum_{i}\lambda_{i}^{llll''}\left\lbrace \Phi_{l}^{\dag}\Phi_{l}^{\dag}\Phi_{l'}\Phi_{l''}\right\rbrace_i+\sum_{i}\lambda_{i}^{ll'll''}\left\lbrace \Phi_{l}^{\dag}\Phi_{l'}^{\dag}\Phi_{l}\Phi_{l''}\right\rbrace_i\\
&&+\sum_{i}\lambda_{i}^{ll'l'l''}\left\lbrace \Phi_{l}^{\dag}\Phi_{l'}^{\dag}\Phi_{l'}\Phi_{l''}\right\rbrace_i+\sum_{i}\lambda_{i}^{l'l'll''}\left\lbrace \Phi_{l'}^{\dag}\Phi_{l'}^{\dag}\Phi_{l}\Phi_{l''}\right\rbrace_i\\
&&+\sum_{i}\lambda_{i}^{l'l'l'l''}\left\lbrace \Phi_{l'}^{\dag}\Phi_{l'}^{\dag}\Phi_{l'}\Phi_{l''}\right\rbrace_i+\sum_{i}\lambda_{i}^{lll''l''}\left\lbrace \Phi_{l}^{\dag}\Phi_{l}^{\dag}\Phi_{l''}\Phi_{l''}\right\rbrace_i\\
&&+\sum_{i}\lambda_{i}^{l'l'l''l''}\left\lbrace \Phi_{l'}^{\dag}\Phi_{l'}^{\dag}\Phi_{l''}\Phi_{l''}\right\rbrace_i+\sum_{i}\lambda_{i}^{ll''ll''}\left\lbrace \Phi_{l}^{\dag}\Phi_{l''}^{\dag}\Phi_{l}\Phi_{l''}\right\rbrace_i\\
&&+\sum_{i}\lambda_{i}^{l'l''l'l''}\left\lbrace \Phi_{l'}^{\dag}\Phi_{l''}^{\dag}\Phi_{l'}\Phi_{l''}\right\rbrace_i\\
 && +\sum_{i}\lambda_{i}^{ll\nu\nu}\left\lbrace \Phi_{l}\Phi_{l}\Phi_{\nu}\Phi_{\nu}\right\rbrace_i+\sum_{i}\lambda_{i}^{ll'\nu\nu}\left\lbrace \Phi_{l}\Phi_{l'}\Phi_{\nu}\Phi_{\nu}\right\rbrace_i\\
&&+\sum_{i}\lambda_{i}^{l'l'\nu\nu}\left\lbrace \Phi_{l'}\Phi_{l'}\Phi_{\nu}\Phi_{\nu}\right\rbrace_i+\sum_{i}\lambda_{i}^{ll''\nu\nu}\left\lbrace \Phi_{l}\Phi_{l''}\Phi_{\nu}\Phi_{\nu}\right\rbrace_i\\
&&+\sum_{i}\lambda_{i}^{l'l''\nu\nu}\left\lbrace \Phi_{l'}\Phi_{l''}\Phi_{\nu}\Phi_{\nu}\right\rbrace_i+\sum_{i}\lambda_{i}^{l''l''\nu\nu}\left\lbrace \Phi_{l''}\Phi_{l''}\Phi_{\nu}\Phi_{\nu}\right\rbrace_i\\
&&+\sum_{i}\lambda_{i}^{ll\nu\nu}\left\lbrace \Phi_{l}^{\dag}\Phi_{l}\Phi_{\nu}^{\dag}\Phi_{\nu}\right\rbrace_i+\sum_{i}\lambda_{i}^{ll'\nu\nu}\left\lbrace \Phi_{l}^{\dag}\Phi_{l'}\Phi_{\nu}^{\dag}\Phi_{\nu}\right\rbrace_i\\
&&+\sum_{i}\lambda_{i}^{l'l'\nu\nu}\left\lbrace \Phi_{l'}^{\dag}\Phi_{l'}\Phi_{\nu}^{\dag}\Phi_{\nu}\right\rbrace_i+\sum_{i}\lambda_{i}^{ll''\nu\nu}\left\lbrace \Phi_{l}^{\dag}\Phi_{l''}\Phi_{\nu}^{\dag}\Phi_{\nu}\right\rbrace_i\\
&&+\sum_{i}\lambda_{i}^{l'l''\nu\nu}\left\lbrace \Phi_{l'}^{\dag}\Phi_{l''}\Phi_{\nu}^{\dag}\Phi_{\nu}\right\rbrace_i+\sum_{i}\lambda_{i}^{ll\nu\nu'}\left\lbrace \Phi_{l}\Phi_{l}\Phi_{\nu}\Phi_{\nu'}\right\rbrace_i\\
&&+\sum_{i}\lambda_{i}^{ll'\nu\nu'}\left\lbrace \Phi_{l}\Phi_{l'}\Phi_{\nu}\Phi_{\nu'}\right\rbrace_i+\sum_{i}\lambda_{i}^{l'l'\nu\nu'}\left\lbrace \Phi_{l'}\Phi_{l'}\Phi_{\nu}\Phi_{\nu'}\right\rbrace_i\\
&&+\sum_{i}\lambda_{i}^{l'l''\nu\nu'}\left\lbrace \Phi_{l'}\Phi_{l''}\Phi_{\nu}\Phi_{\nu'}\right\rbrace_i+\sum_{i}\lambda_{i}^{ll\nu\nu'}\left\lbrace \Phi_{l}^{\dag}\Phi_{l}\Phi_{\nu}^{\dag}\Phi_{\nu'}\right\rbrace_i\\
&&+\sum_{i}\lambda_{i}^{ll'\nu\nu'}\left\lbrace \Phi_{l}^{\dag}\Phi_{l'}\Phi_{\nu}^{\dag}\Phi_{\nu'}\right\rbrace_i+\sum_{i}\lambda_{i}^{l'l\nu\nu'}\left\lbrace \Phi_{l'}^{\dag}\Phi_{l}\Phi_{\nu}^{\dag}\Phi_{\nu'}\right\rbrace_i\\
&&+\sum_{i}\lambda_{i}^{l'l'\nu\nu'}\left\lbrace \Phi_{l'}^{\dag}\Phi_{l'}\Phi_{\nu}^{\dag}\Phi_{\nu'}\right\rbrace_i+\sum_{i}\lambda_{i}^{ll''\nu\nu'}\left\lbrace \Phi_{l}^{\dag}\Phi_{l''}\Phi_{\nu}^{\dag}\Phi_{\nu'}\right\rbrace_i\\
&&+\sum_{i}\lambda_{i}^{\nu\nu\nu\nu'}\left\lbrace \Phi_{\nu}^{\dag}\Phi_{\nu}\Phi_{\nu}^{\dag}\Phi_{\nu'}\right\rbrace_i+\sum_{i}\lambda_{i}^{\nu\nu'\nu\nu'}\left\lbrace \Phi_{\nu}^{\dag}\Phi_{\nu'}\Phi_{\nu}^{\dag}\Phi_{\nu'}\right\rbrace_i\\
&&+\sum_{i}\lambda_{i}^{\nu\nu\nu'\nu'}\left\lbrace \Phi_{\nu}^{\dag}\Phi_{\nu}\Phi_{\nu'}^{\dag}\Phi_{\nu'}\right\rbrace_i\\
\end{eqnarray*}

\begin{eqnarray*}
&&+\sum_{i}\lambda_{i}^{ll\nu'\nu'}\left\lbrace \Phi_{l}\Phi_{l}\Phi_{\nu'}\Phi_{\nu'}\right\rbrace_i+\sum_{i}\lambda_{i}^{l'l'\nu'\nu'}\left\lbrace \Phi_{l'}\Phi_{l'}\Phi_{\nu'}\Phi_{\nu'}\right\rbrace_i\\
&&+\sum_{i}\lambda_{i}^{l''l''\nu'\nu'}\left\lbrace \Phi_{l''}\Phi_{l''}\Phi_{\nu'}\Phi_{\nu'}\right\rbrace_i+\sum_{i}\lambda_{i}^{ll''\nu\nu'}\left\lbrace \Phi_{l}^{\dag}\Phi_{l''}\Phi_{\nu'}^{\dag}\Phi_{\nu}\right\rbrace_i\\
&&+\sum_{i}\lambda_{i}^{ll\nu'\nu'}\left\lbrace \Phi_{l}^{\dag}\Phi_{l}\Phi_{\nu'}^{\dag}\Phi_{\nu'}\right\rbrace_i+\sum_{i}\lambda_{i}^{l'l'\nu'\nu'}\left\lbrace \Phi_{l'}^{\dag}\Phi_{l'}\Phi_{\nu'}^{\dag}\Phi_{\nu'}\right\rbrace_i\\
&&+\sum_{i}\lambda_{i}^{l''l''\nu\nu}\left\lbrace \Phi_{l''}^{\dag}\Phi_{l''}\Phi_{\nu}^{\dag}\Phi_{\nu}\right\rbrace_i+\sum_{i}\lambda_{i}^{l''l''\nu'\nu'}\left\lbrace \Phi_{l''}^{\dag}\Phi_{l''}\Phi_{\nu'}^{\dag}\Phi_{\nu'}\right\rbrace_i\\
&&+h.c., \\
\end{eqnarray*}

where $\sum_i\lambda_i\left\lbrace \right\rbrace_i$, $\sum_i\kappa_i\left\lbrace \right\rbrace_i$ sums over all possible ways to group the fields inside the brackets and make the product of representations in order to obtain a singlet.

\bibliographystyle{h-physrev4}


\end{document}